\documentclass[aps,prd,preprint,groupedaddress,showpacs]{revtex4}
\usepackage{epsfig}
\usepackage{graphics}
\usepackage{slashed}
\usepackage{color}
\begin{document}

\leftmargin -2cm
\def\choosen{\atopwithdelims..}




\title{Prompt $J/\psi$ production in the Regge
limit of QCD: \\ From Tevatron to LHC}

\author{\firstname{V.A.} \surname{Saleev}}
\email{saleev@ssu.samara.ru, saleev@mail.desy.de}

 \affiliation{{II.} Institut f\"ur Theoretische Physik, Universit\" at Hamburg,
Luruper Chaussee 149, 22761 Hamburg, Germany} \affiliation{ Samara
State University, Academic Pavlov Street 1, 443011 Samara, Russia}


\author{\firstname{M.A.
}\surname{Nefedov}}\email{nefedovma@gmail.com}
\author{\firstname{A.V. }\surname{Shipilova}}\email{alexshipilova@ssu.samara.ru}
\affiliation{ Institut f. Kernphysik, Forschungszentrum Juelich,
52425 Juelich, Germany} \affiliation{ Samara State University,
Academic Pavlov Street 1, 443011 Samara, Russia}

\begin{abstract}
We study prompt $J/\psi-$meson hadroproduction invoking the
hypothesis of gluon Reggeization in $t-$channel exchanges at high
energy and the factorization formalism of nonrelativistic quantum
chromodynamics at the leading order in the strong-coupling constant
$\alpha_s$ and the relative velocity of quarks $v$. The
transverse-momentum distribution of direct and prompt $J/\psi-$meson
production measured at the Fermilab Tevatron fitted to obtain the
nonperturbative long-distance matrix elements, which are used to
predict prompt $J/\psi$ production spectra at the CERN LHC. At the
numerical calculation, we adopt the Kimber-Martin-Ryskin and
Bl\"umlein prescriptions to derive unintegrated gluon distribution
function of the proton from their collinear counterpart, for which
we use the Martin-Roberts-Stirling-Thorne set. Without adjusting any
free parameters, we find good agreement with measurements by the
ATLAS, CMS and LHCb Collaborations at the LHC at the hadronic c.m.\
energy $\sqrt S=7$ TeV.
\end{abstract}

\maketitle

\section{Introduction}
\label{sec:one}

The production of heavy quarkonium at hadron colliders provides
useful laboratory for testing the high-energy limit of quantum
chromodynamics (QCD) as well as the interplay of perturbative and
nonperturbative phenomena in QCD.

The total collision energies, $\sqrt{S}=1.8$~TeV and 1.96~TeV in
Tevatron Runs~I and II, respectively, and $\sqrt{S}=7$~TeV or
14~TeV at the LHC, sufficiently exceed the characteristic scale
$\mu$ of the relevant hard processes, which is of order of
quarkonium transverse mass $M_T=\sqrt{M^2+p_T^2}$, {\it i.e.}\ we
have $\Lambda_\mathrm{QCD}\ll\mu\ll\sqrt{S}$. In this high-energy
regime, so called "Regge limit", the contribution of partonic
subprocesses involving $t$-channel parton (gluon or quark)
exchanges to the production cross section can become dominant.
Then the transverse momenta of the incoming partons and their
off-shell properties can no longer be neglected, and we deal with
"Reggeized" $t$-channel partons. These $t-$channel exchanges obey
multi-Regge kinematics (MRK), when the particles produced in the
collision are strongly separated in rapidity. If the same
situation is realized with groups of particles, then
quasi-multi-Regge kinematics (QMRK) is at work. In the case of
$J/\psi-$meson inclusive production, this means the following:
$J/\psi-$meson (MRK) or $J/\psi$-meson plus gluon jet (QMRK) is
produced in the central region of rapidity, while other particles
are produced with large modula of rapidities.

The parton Reggeization approach (PRA) \cite{QMRK,Lipatov95} is
particularly appropriate for high-energy phenomenology. We see,
the assumption of a dominant role of MRK or QMRK production
mechanisms at high energy works well. PRA is based on an effective
quantum field theory implemented with the non-Abelian
gauge-invariant action including fields of Reggeized gluons
\cite{BFKL} and Reggeized quarks \cite{LipatoVyazovsky}. Reggeized
partons interact with quarks and Yang-Mills gluons in a specific
way. Recently, in Ref.\cite{Antonov}, the Feynman rules for the
effective theory of Reggeized gluons were derived for the induced
and some important effective vertices. This approach was
successfully applied to interpret the production of isolated jets
\cite{KSS2011}, prompt photons \cite{SVADISy}, diphotons
\cite{SVAdiy}, charmed mesons \cite{PRD}, bottom-flavored jets
\cite{PRb} measured at the Fermilab Tevatron, at the DESY HERA and
at the CERN LHC, in the small-$p_T$ regime, where $p_T<<\sqrt{S}$.
We suggest the MRK (QMRK) production mechanism to be the dominant
one at small $p_T$ values. Using the Feynman rules for the
effective theory, we can construct heavy quarkonium production
amplitudes in framework of non-relativistic QCD
(NRQCD)\cite{NRQCD,Maltoni}.

The factorization formalism of the NRQCD  is a rigorous theoretical
framework for the description of heavy-quarkonium production and
decay. The factorization hypothesis of NRQCD assumes the separation
of the effects of long and short distances in heavy-quarkonium
production. NRQCD is organized as a perturbative expansion in two
small parameters, the strong-coupling constant $\alpha_s$ and the
relative velocity $v$ of the heavy quarks inside a heavy quarkonium.

Our previous analysis of charmonium \cite{KSVcharm, SVpepan} and
bottomonium \cite{KSVbottom,SVpepan} production at the Fermilab
Tevatron using the high-energy factorization scheme and NRQCD
approach has shown the efficiency of such type of high-energy
phenomenology. In this paper we repeat our calculations for the
direct and prompt $J/\psi-$meson transverse momentum spectra at
the Fermilab Tevatron \cite{KSVcharm} to obtain by fitting
procedure octet nonperturbative matrix elements (NMEs), than we
calculate prompt $J/\psi-$meson spectra, which were measured
recently at the CERN LHC Collider at the energy of  $\sqrt S=7$
TeV. We find a good agreement of our calculations and experimental
data from ATLAS \cite{ATLASpsi}, CMS~\cite{CMSpsi} and LHCb
\cite{LHCbpsi} Collaborations.

\section{Model}

Working at the leading order (LO) in  $\alpha_s$ and $v$ we consider
the following partonic subprocesses, which describe charmonium
production at high energy:
\begin{eqnarray}
R(q_1) + R(q_2) &\to& {\cal H}
[{^3P}_J^{(1)},{^3S}_1^{(8)},{^1S}_0^{(8)},{^3P}_J^{(8)}](p),
\label{eq:RRtoH}\\
 R(q_1) + R(q_2) &\to& {\cal H} [{^3S}_1^{(1)}](p) + g(p'),
\label{eq:RRtoHG}
\end{eqnarray}
where $R$ is the Reggeeized gluon and $g$ is the Yang-Mills gluon,
respectively, with four-momenta indicated in parentheses, ${\cal
H}[n]$ is the physical charmonium state, $n={}^{2S+1}L_J^{(1,8)}$ is
the included $c\bar c$ Fock state with the spin $S$, total angular
momentum $J$, orbital angular momentum $L$ and in the singlet
$^{(1)}$ or in the octet $^{(8)}$ color state.

In the general case, the partonic cross section of charmonium
production receives from the $c\bar c$ Fock state
$[n]=[{}^{2S+1}L_J^{(1,8)}]$ the contribution \cite{NRQCD,Maltoni}
\begin{equation}
d\hat \sigma (R + R \to c\bar c[{}^{2S+1}L_J^{(1,8)}] \to {\cal
H})=d\hat \sigma (R + R \to c\bar
c[^{2S+1}L_J^{(1,8)}])\frac{\langle {\cal O}^{\cal
H}[^{2S+1}L_J^{(1,8)}]\rangle}{N_\mathrm{col}N_\mathrm{pol}},
\end{equation}
where $N_\mathrm{col}=2 N_c$ for the color-singlet state,
$N_\mathrm{col}=N_c^2-1$ for the color-octet state, and
$N_\mathrm{pol}=2J+1$, $\langle {\cal O}^{\cal
H}[^{2S+1}L_J^{(1,8)}]\rangle$ are the NMEs. They satisfy the
multiplicity relations
\begin{eqnarray}
\langle{\cal O}^{\psi(nS)}[^3P_J^{(8)}]\rangle&=&(2J+1)\langle{\cal
O}^{\psi(nS)}[^3P_0^{(8)}]\rangle,\nonumber\\
\langle{\cal O}^{\chi_{cJ}}[^3P_J^{(1)}]\rangle&=&(2J+1)\langle{\cal
O}^{\chi_{c0}}[^3P_0^{(1)}]\rangle,\nonumber\\
\langle{\cal O}^{\chi_{cJ}}[^3S_1^{(8)}]\rangle&=&(2J+1)\langle{\cal
O}^{\chi_{c0}}[^3S_1^{(8)}]\rangle,
\end{eqnarray}
which follow from heavy-quark spin symmetry to LO in $v$.

The partonic cross section of $c\bar c$ production is defined as
\begin{equation}
d\hat\sigma(R + R \to c\bar
c[^{2S+1}L_J^{(1,8)}])=\frac{1}{I}\overline{|{\cal A}(R + R \to
c\bar c[^{2S+1}L_J^{(1,8)}])|^2}d\Phi,\label{eq:dsigma}
\end{equation}
where $I=2 x_1 x_2 S$ is the flux factor of the incoming particles,
which is taken as in the collinear parton model \cite{KTCollins},
${\cal A}(R + R \to c\bar c[{}^{2S+1}L_J^{(1,8)}])$ is the
production amplitude, the bar indicates average (summation) over
initial-state (final-state) spins and colors, and $d\Phi$ is the
invariant phase space volume of the outgoing particles. This
convention implies that the cross section in the high-energy
factorization scheme is normalized approximately to the cross
section for on-shell gluons in the collinear parton model when ${\bf
q}_{1T}={\bf q}_{2T}={\bf 0}$.

The LO results for the squared amplitudes of
subprocesses~(\ref{eq:RRtoH}) and (\ref{eq:RRtoHG}) that we found
by using the Feynman rules of Ref.~\cite{Antonov} coincide with
those we obtained in Ref.~\cite{KSVcharm}. The formulas for the
squared amplitudes $\overline{|{\cal A}(R + R \to c\bar
c[^{2S+1}L_J^{(1,8)}])|^2}$ for the $2\to 1$ subprocesses
(\ref{eq:RRtoH}) are listed in Eq. (27) of Ref.~\cite{KSVcharm}.
The analytical result in case of the $2\to 2$ subprocess
(\ref{eq:RRtoHG}) is presented in Ref.\cite{SVpepan}, where the
results for the $2\to 1$ subprocesses are also listed, but in
another equivalent form. The relation between these forms is
discussed in Ref.\cite{KSVbottom}.

Exploiting the hypothesis of high-energy factorization, we may write
the hadronic cross section $d\sigma$ as convolution of partonic
cross section $d\hat \sigma$ with unintegrated parton distribution
functions (PDFs) $\Phi_g^p(x,t,\mu^2)$ of Reggeized gluon in the
proton, as
\begin{eqnarray}
d\sigma(p + p \to {\cal H} + X)&=&  \int\frac{d x_1}{x_1}
\int\frac{d^2{\bf q}_{1T}}{\pi} \Phi_g^p\left(x_1,t_1,\mu^2\right)
\int\frac{d x_2}{x_2} \int\frac{d^2 {\bf q}_{2T}}{\pi}
\nonumber\\
&&{}\times\Phi_g^p\left(x_2,t_2,\mu^2\right) d\hat\sigma(R + R \to
{\cal H}+X). \label{eq:KT}
\end{eqnarray}
$t_1=|{\bf q}_{1T}|^2$, $t_2=|{\bf q}_{2T}|^2$, $x_1$ and $x_2$
are the fractions of the proton momenta passed on to the Reggeized
gluons, and the factorization scale $\mu$ is chosen to be of order
$M_T$. The collinear and unintegrated gluon distribution functions
are formally related as
\begin{equation}
xG^p(x,\mu^2)=\int^{\mu^2} \Phi_g^p(x,t,\mu^2)dt, \label{eq:xG}
\end{equation}
so that, for ${\bf q}_{1T}={\bf q}_{2T}={\bf 0}$, we recover the
conventional factorization formula of the collinear parton model,
\begin{equation}
d\sigma(p + p \to {\cal H}+\!X)=\int{d x_1}G^p(x_1,\mu^2) \int{d
x_2} G^p(x_2,\mu^2) d\hat \sigma(g + g \to {\cal H} + X).
\label{eq:PM}
\end{equation}

We now describe how to evaluate the differential hadronic cross
section from Eq.~(\ref{eq:KT}) combined with the squared amplitudes
of the $2\to1$ and $2\to2$ subprocesses~(\ref{eq:RRtoH}) and
(\ref{eq:RRtoHG}), respectively. The rapidity and pseudorapidity of
a charmonium state with four-momentum $p^\mu=(p^0,{\bf p}_{T},p^3)$
are given by
\begin{equation}
y=\frac{1}{2}\ln\frac{p^0+p^3}{p^0-p^3},\quad
\eta=\frac{1}{2}\ln\frac{|{\bf p}|+p^3}{|{\bf p}|-p^3},
\end{equation}
respectively, and $dy=\displaystyle\frac{|{\bf p}|}{p^0}d\eta$.


 The invariant phase volume $d\Phi$ in the
Eq.~(\ref{eq:dsigma}) for $2\to 1$ subprocess (\ref{eq:RRtoH}) can
be presented as follows:
\begin{eqnarray}
d\Phi({\bf p})&=&(2\pi)^4\delta^{(4)}(q_1+q_2-p)\frac{d^3p}{(2\pi)^3
2p^0}\nonumber\\
&=&\frac{4\pi^2
p_T}{S}\delta(\xi_1-\frac{p^0+p^3}{\sqrt{S}})\delta(\xi_2-\frac{p^0-p^3}{\sqrt{S}})
\delta^{2}({\bf q}_{1T}+{\bf q}_{2T}-{\bf
p}_{T})dp_Tdy.\label{eq:21}
\end{eqnarray}
From the Eqs. (\ref{eq:dsigma}), (\ref{eq:KT}) and (\ref{eq:21}) we
obtain the master formula for the $2\to1$
subprocess~(\ref{eq:RRtoH}):
\begin{eqnarray}
&&\frac{d\sigma(p + p \to {\cal H} + X)} {d p_Td y} = \frac{p_T}{(
p_T^2+M^2)^2} \int{dt_1}\int{d \varphi_1}
\nonumber\\
&&{}\times \Phi_g^p(\xi_1,t_1,\mu^2) \Phi_g^p(\xi_2,t_2,\mu^2)
\overline{|{\cal A}(R + R \to {\cal H})|^2},
\end{eqnarray}
where $t_2=t_1+p_T^2-2 p_T \sqrt{t_1}\cos (\phi_1)$ and the relation
$\xi_1\xi_2 S=p_T^2+M^2$ has been taken into account.

The invariant phase volume $d\Phi$ in the Eq.~(\ref{eq:dsigma}) for
$2\to 2$ subprocess (\ref{eq:RRtoHG}) can be presented as follows:
\begin{eqnarray}
d\Phi({\bf p},{\bf
p'})&=&(2\pi)^4\delta^{(4)}(q_1+q_2-p-p')\frac{d^3p}{(2\pi)^3
2p^0}\frac{d^3{p}'}{(2\pi)^3
2{p}'^0}\nonumber\\
&=&\frac{p_T}{4\pi}\delta((q_1+q_2-p)^2)dp_Tdy.\label{eq:22}
\end{eqnarray}
Such a way, accordingly the Eqs. (\ref{eq:dsigma}), (\ref{eq:KT})
and (\ref{eq:22}), we have the master formula for the $2\to2$
subprocess~(\ref{eq:RRtoHG}):
\begin{eqnarray}
&&\frac{d\sigma(p + p \to {\cal H} + X)}{d{p_T d y}}=\frac{p_T}{(2
\pi)^3} \int{dt_1}\int{d \varphi_1} \int{d x_2}\int{dt_2}\int{d
\varphi_2}\nonumber\\&&{} \times \Phi_g^p(x_1,t_1,\mu^2)
\Phi_{g}^p(x_2,t_2,\mu^2) \frac{\overline{|{\cal A}(R + R \to {\cal
H} + g)|^2}}{ (x_2 - \xi_2)(2 x_1 x_2 S)^2},
\end{eqnarray}
where $\phi_{1,2}$ are the angles enclosed between $\vec {\bf
q}_{1,2T}$ and the transverse momentum $\vec {\bf p}_T$ of ${\cal
H}$,
\begin{equation}
x_1=\frac{1}{(x_2 - \xi_2) S}\left[({\bf q}_{1T}+{\bf q}_{2T} - {\bf
p}_T)^2- M^2 - |{\bf p}_{T}|^2 + x_2 \xi_1 S\right].
\end{equation}


In our numerical analysis, we adopt as our default the prescription
proposed by Kimber, Martin, and Ryskin (KMR) \cite{KMR} to obtain
unintegrated gluon PDF of the proton from the conventional
integrated one, as implemented in Watt's code \cite{Watt}. As is
well known \cite{Andersson:2002cf}, other popular prescriptions,
such as those by Bl\"umlein \cite{Blumlein:1995eu} or by Jung and
Salam \cite{Jung:2000hk}, produce unintegrated PDFs with distinctly
different $t$ dependences. In order assess the resulting theoretical
uncertainty, we also evaluate the unintegrated gluon PDF using the
Bl\"umlein approach, which resums small-$x$ effects according to the
Balitsky-Fadin-Kuraev-Lipatov (BFKL) equation \cite{BFKL}. As input
for these procedures, we use the LO set of the
Martin-Roberts-Stirling-Thorne (MRST) \cite{MRST98,MRST2002} proton
PDF as our default.

Throughout our analysis the renormalization and factorization scales
are identified and chosen to be $\mu=\xi M_T$, where $\xi$ is varied
between 1/2 and 2 about its default value 1 to estimate the
theoretical uncertainty due to the freedom in the choice of scales.
The resulting errors are indicated as shaded bands in the figures.

\section{Results}

At first, we examine direct and prompt (sum of direct production,
production via radiative decays of $\chi_{cJ}$ mesons and production
via decays of $\psi'$ mesons contributions) $J/\psi-$meson
production in proton-antiproton collisions at the Tevatron at the
energies $\sqrt{S}=1.8$~TeV (Run~I)\cite{CDFI} and
$\sqrt{S}=1.96$~TeV (Run~II)\cite{CDFII} in the central region of
pseudorapidity $|\eta|<0.6$. The data of Run~II includes $p_T$
distributions of prompt $J/\psi$ mesons, so far without separation
into direct, $\chi_{cJ}-$decay, and $\psi'-$decay contributions. In
the Ref.\cite{KSVcharm}, we have performed a joint fit to the Run~I
and Run~II CDF data \cite{CDFI,CDFII} to obtain the color-octet
NME's for $J/\psi$, $\psi'$, and $\chi_{cJ}$ mesons. Our fits
included five experimental data sets, which come as $p_T$
distributions of $J/\psi$ mesons from direct production, prompt
production, $\chi_{cJ}$ decays, and $\psi'$ decays. Since in the
previous calculations we have used the old set MRST98 \cite{MRST98}
as a collinear input for unintegrated PDF, in the present study we
repeat our fit with the next-generation MRST set \cite{MRST2002}. We
find small differences between the old and new fit parameters,
however, it is important for precise description of the data. The
results of our fit are presented in the Table~I along with results
of the fit in the next to leading order (NLO) of collinear parton
model (PM) and NRQCD approach~\cite{KniehlPSI}. Oppositely the
Ref.\cite{KniehlPSI}, we perform a fit procedure by assumption for
NMEs to be only positive. Than, using the CDF data for {a} prompt
$J/\psi$ production \cite{CDFI}, presented separately for direct
$J/\psi$ mesons, $J/\psi$ from $\psi^\prime$ decays, and $J/\psi$
from $\chi_{cJ}$ decays, we obtain color-octet NMEs $\langle {\cal
O}^{J/\psi}[^3S_1^{(8)}, ^1S_0^{(8)},^3P_0^{(8)}]\rangle$, $\langle
{\cal O}^{\psi^\prime}[^3S_1^{(8)},
^1S_0^{(8)},^3P_0^{(8)}]\rangle$, and $\langle {\cal
O}^{\chi_{c0}}[^3S_1^{(8)}]\rangle$ independently from each other.

Looking at the Table I, we {find} a good agreement with the NLO fit
in collinear parton model performed in the Ref.~\cite{KniehlPSI},
which strongly improves if we take into account that a sum of
contributions of NMEs $\langle {\cal
O}^{J/\psi}[^1S_0^{(8)}]\rangle$ and $\langle {\cal
O}^{J/\psi}[^3P_0^{(8)}]\rangle$ from the Ref.~\cite{KniehlPSI},
leading to almost parallel $J/\psi$ transverse momenta spectra,
corresponds to our contribution of the NME $\langle {\cal
O}^{J/\psi}[^1S_0^{(8)}]\rangle$. Such an agreement demonstrates a
validity of factorization in the charmonium production in hadronic
collisions, i.e. an independence of the $c\bar c$ production
mechanism from the nonperturbative charmonium formation at the last
step. {It is necessary to note that a same consent between LO
results obtained in the uncollinear factorization scheme and NLO
results obtained in the collinear parton model is also observed when
describing other relevant processes, see
Refs.~\cite{KSS2011,SVADISy,SVAdiy,PRD,PRb}.}

In Figs.~1--4, we compare the CDF  data on $J/\psi$ mesons from
direct production, $\psi'$ decays and $\chi_{cJ}$ decays in Run~I
\cite{CDFI} and from prompt production in Run~II \cite{CDFII}, with
the respective theoretical results evaluated with the NMEs listed in
Table~I. As default, we present in all figures the theoretical
results which are obtained using KMR unintegrated gluon density
\cite{KMR}. The comparison between KMR~\cite{KMR} and
Bl\"umlein~\cite{Blumlein:1995eu} PDFs is made in Fig.~4 for a
prompt $J/\psi$ production only.

In Fig.~1 one can find a dominance of color-octet contributions at
all values of direct $J/\psi$ meson transverse momentum:
$[{}^3S_1^{(8)}]$ contribution dominates at the large values
$p_T>10$ GeV, and $[{}^1S_0^{(8)}]$
--- at the small $p_T<10$ GeV. The situation is very similar for
$J/\psi$ production from $\psi'$ decay, considered in Fig.~2. It is
also important, that the $[{}^3P_J^{(8)}]$ contribution vanishes in
direct $J/\psi$ and $\psi'$ production. The obtained results are in
agreement with previous calculations of Ref.~\cite{KSVcharm} with a
slight difference. In Ref.~\cite{Teryaev1} it was also shown that in
the direct $J/\psi$ production at the Tevatron color-octet
contribution dominates. Oppositely our conclusions, in
Ref.~\cite{Teryaev1}, the main contribution comes from the
$[{}^1S_0^{(8)}]$ NME at all transverse momenta. The reason of this
discrepancy arises from the fact that in Ref.~\cite{Teryaev1} the
color-octet NMEs for $J/\psi$ meson have been obtained by a fit of
direct $J/\psi$ production data for $p_T>5$
 GeV only. In our fit we take into account both direct \cite{CDFI} and
 prompt \cite{CDFII}
 $J/\psi$ production data, the last ones contain points in a small $p_T$
 region. We observe, the inclusion of prompt $J/\psi$ production
 data in the fit to change the relative weight of different color-octet NMEs.
 This fact should be important in study of polarized $J/\psi$
 mesons.

In case of $J/\psi$ production via decays of $\chi_{cJ}$ mesons,
considered in Fig.~3, we confirm the conclusion of
Refs.\cite{Teryaev2,KSVcharm}, which reads, that in the
high-energy factorization scheme, the color-singlet contribution
is sufficient to describe the data for production of $P-$wave
charmonium states.

In Fig.~4, the $p_T$ distribution of prompt $J/\psi$ production in
Tevatron Run~II is presented as a sum of the contributions from
direct production, $\psi'$ decays, and $\chi_{cJ}$ decays. We
observe at the $p_T\geq 2$ GeV the $J/\psi$ mesons to be produced
preferably directly. The contribution from $\chi_{cJ}$ decays
dominates at small $p_T<2$ GeV. The contribution from $\psi'$ is
smaller than other ones and it exceeds the contribution from
$\chi_{cJ}$ decays only at the $p_T>16$ GeV.

The curve number (5) in Fig.~4 is obtained using Bl\"umlein
unPDF~\cite{Blumlein:1995eu}. The visible difference between this
curve and the curve (4), which is obtained using KMR unPDF
\cite{KMR}, takes place only in the region of small $J/\psi$
transverse momentum. In the range of $p_T\geq 5$ GeV, there is no
difference between them as in the prompt production as in the direct
production or in the production via decays of high charmonium
states, $\psi'$ and $\chi_{cJ}$.

As it is obvious from Figs.~1--4, the theoretical uncertainties
associated with the variation of the factorization scale $\mu$ are
large at the small $p_T$ region, taking a value of about factor 5
between upper and lower boundaries, and they sufficiently decrease
down to a factor 2 at the $p_T\geq 6$~GeV. The uncertainties from
errors in the color-octet NMEs are small, they are about 7-10\%.

Moving on from Tevatron to the LHC, which is currently running at
the total energy being about 3.5 times larger than at the
Tevatron, we expect the range of validity of our approach to be
extended by the same factor, to $p_T\leq 70$ GeV, as we describe
well the Tevatron data at the range of $0<p_T<20$ GeV. This
expectation is nicely confirmed in Figs.~5--6, where the recent
measurements of the prompt $J/\psi$ production by the ATLAS
Collaboration at the CERN LHC \cite{ATLASpsi}, which cover the
kinematic region 1 GeV $<p_T<70$ GeV and $|y|<2.4$, are compared
with our predictions based on the particle Reggeization approach
and NRQCD formalism. The measurements of the CMS Collaboration
\cite{CMSpsi} were performed in the similar kinematic range 6.5
GeV $<p_T< 30$ GeV and $|y|<2.4$, see Fig.~7. We observe a
dominant role of direct production mechanism in the prompt
$J/\psi$ hadroproduction at the all values of $J/\psi$ meson
transverse momentum. Concerning the relative contributions of
$\psi'$ decays and $\chi_{cJ}$ decays into a prompt $J/\psi$
production, we found the contribution from $\psi'$ decays to
dominate at the large $p_T>20$ GeV, and the contribution from
$\chi_{cJ}$ decays to dominate at the small $p_T$, respectively.
Additionally, we compare our predictions with the data from LHCb
Collaboration \cite{LHCbpsi}, which were extracted in the range
$0<p_T<14$ GeV and $2<|y|<4.5$. We find a good agreement between
our predictions and prompt $J/\psi$ production data at the
moderate rapidity interval $2.0<|y|<3.5$, see Figs.~8--9. At the
same time our theoretical result overestimates the data of at most
factor 2 in the range of large rapidity $3.5<|y|<4.5$. This
distinction is expected in the parton Reggeization approach,
because the multi-Regge kinematics conditions to be broken if
$J/\psi$ mesons are produced with large rapidity.

We observe, in Fig.~10, that relative contributions of the
color-singlet (curve 1) and color-octet (curve 2) production
mechanisms to the prompt $J/\psi$  spectrum strongly depend on the
$J/\psi$ transverse momentum. Similarly to the NLO calculations in
the collinear parton model, the color-octet contribution dominates
at the large $p_T$ region, basically via the contributions of the
color-octet NMEs $\langle {\cal O}^{J/\psi}[^3S_1^{(8)}]\rangle$ and
$\langle {\cal O}^{\psi^\prime}[^3S_1^{(8)}]\rangle$. It is
significant, the experimental data \cite{ATLASpsi,CMSpsi,LHCbpsi}
depend on the assumption of polarization of produced $J/\psi$ mesons
slightly. We perform calculations and make a comparison to the data
in a case of non-polarized $J/\psi$ meson production.

Comparing our results with the recent studies of $J/\psi$ meson
hadroproduction in the conventional collinear PM, which were
performed in full NLO approximation of NRQCD formalism
\cite{KniehlPSI} or in the non-complete NNLO$^{*}$ approximation
of color-singlet model \cite{Lansberg}, we should emphasize the
following. {At first, oppositely to NLO and NNLO$^{*}$
calculations, which provide a good description of data only at
non-small $p_T>5$ GeV, we can reproduce data well at all
transverse momenta $p_T$.} {At second, the present study along
with the previous investigations in the parton Reggeization
approach
\cite{KSS2011,SVADISy,SVAdiy,PRD,PRb,KSVcharm,SVpepan,KSVbottom,Teryaev1,Teryaev2}
demonstrate the {important} role of (quasi)multi-Regge kinematics
in particle production at high energies, this feature is out of
account in the collinear PM.} {Such a way, we find the approach
based on the effective theory of Reggeized partons
\cite{BFKL,Lipatov95} and high-energy factorization scheme with
unintegrated PDFs, which in the large logarithmic terms
($\ln(\mu^2/\Lambda_{QCD}^2),~ \ln(S/\mu^2)$) are resummed in all
orders of strong coupling constant $\alpha_s$, to be more adequate
for the description of experimental data than fixed order in
$\alpha_s$ calculations in the frameworks of collinear PM.}

\section{Conclusions}
\label{sec:five}

The Fermilab Tevatron and, even more so, the CERN LHC are currently
probing particle physics at terascale c.m.\ energies $\sqrt{S}$, so
that the hierarchy $\Lambda_\mathrm{QCD}\ll\mu\ll\sqrt{S}$, which
defines the MRK and QMRK regimes, is satisfied for processes of
heavy quark and heavy quarkonium production in the central region of
rapidity, where $\mu$ is of order of their transverse mass. In this
paper, we studied  QCD processes of particular interest, namely
prompt $J/\psi$ hadroproduction, at LOs in the parton Reggeization
approach and NRQCD approach, in which they are mediated by $2\to1$
and $2\to 2$ partonic subprocesses initiated by Reggeized gluon
collisions.

We found by the fit of Tevatron data that numerical values of the
color-octet NMEs are very similar to ones obtained in the full NLO
calculations based on NRQCD approach \cite{KniehlPSI}. Using these
NMEs, we nicely described recent LHC data for prompt $J/\psi$ meson
production measured by ATLAS~\cite{ATLASpsi}, CMS~\cite{CMSpsi} and
LHCb~\cite{LHCbpsi} Collaborations at the whole presented range of
$J/\psi$ transverse momenta. We found only one exclusion, the region
of large modulo of rapidity $|y|>3.5$, where LHCb data are by a
factor 2 smaller than our predictions. However, this kinematical
region is out of the applicability limits of the MRK or QMRK
pictures. Here and in
Refs.~\cite{KSS2011,SVADISy,SVAdiy,PRD,PRb,KSVcharm,SVpepan,KSVbottom,Teryaev1,Teryaev2},
parton Reggeization approach was demonstrated to be a powerful tool
for the theoretical description of QCD processes in the high-energy
limit.

\section{Acknowledgements}

We are grateful to B.~A.~Kniehl, M.~Butensch\"{o}n, M.~B\"{u}scher
and N.~N.~Nikolaev for useful discussions. The work of M.~A.~N. and
A.~V.~S. was supported by the Federal Ministry for Science and
Education of the Russian Federation under Contract
No.~14.740.11.0894 and in part by the Grant DFG 436 RUS 113/940; the
work of V.~A.~S. was supported in part by the Russian Foundation for
Basic Research under Grant 11-02-00769-a and  by SFB Fellowship of
Hamburg University (SFB-676).

\newpage
\begin{table}[hp]
\begin{center}
\begin{ruledtabular}

\caption{\label{tab:NME} NMEs for $J/\psi$, $\psi^\prime$, and
$\chi_{cJ}$ mesons from fits of the CDF data~\cite{CDFI,CDFII} in
the NLO collinear parton model \cite{KniehlPSI} and in the parton
Reggeization approach using the Bl\"umlein \cite{Blumlein:1995eu},
and KMR \cite{KMR} unintegrated gluon distribution functions. The
errors on our fit results are determined by the varying in turn each
NME up and down about its central value until the value of $\chi^2$
is increased by unity keeping all other NMEs fixed at their central
values. When we obtained the value of $\chi^2/\text{d.o.f.}
>1$, we have used normalizing multiplier approach \cite{brand} to reduce this
value to unity.}

\begin{tabular}{cccc}
NME  & PM NLO \cite{KniehlPSI} & Fit B & Fit KMR \\
\hline $\langle {\cal O}^{J/\psi}[^3S_1^{(1)}]\rangle/$GeV$^3$ & 1.3 & 1.3 & 1.3 \\
$\langle {\cal O}^{J/\psi}[^3S_1^{(8)}]\rangle/$GeV$^3$ & $(1.68\pm
0.46)\times10^{-3}$ & $(1.89\pm 0.27)\times10^{-3}$ &
$(2.23\pm 0.27)\times10^{-3}$ \\
$\langle{\cal O}^{J/\psi}[^1S_0^{(8)}]\rangle/$GeV$^3$  & $(3.04\pm
0.35)\times10^{-2}$ & $(1.80\pm 0.25)\times10^{-2}$ &
$(1.84\pm 0.19)\times10^{-2}$ \\
$\langle {\cal O}^{J/\psi}[^3P_0^{(8)}]\rangle/$GeV$^5$ &$ (-9.08\pm
1.61)\times10^{-3}$ & 0 & 0
\\
\hline
$\chi^2/\mathrm{d.o.f}$   & --- & 1.0 & 1.0 \\
\hline $\langle {\cal O}^{\psi^\prime}[^3S_1^{(1)}]\rangle/$GeV$^3$
 & $6.5\times10^{-1}$ & $6.5\times10^{-1}$ &
$6.5\times10^{-1}$ \\
$\langle {\cal O}^{\psi^\prime}[^3S_1^{(8)}]\rangle/$GeV$^3$ &
$(1.88\pm0.62)\times10^{-3}$ & $(6.72\pm 1.15)\times 10^{-4}$ &
$(9.33\pm 1.62)\times10^{-4}$ \\
$\langle{\cal O}^{\psi^\prime}[^1S_0^{(8)}]\rangle/$GeV$^3$ & $(7.01\pm 4.75)
\times 10^{-3}$ & $(3.63\pm 1.40)\times10^{-3}$ & $(3.27\pm 1.44)\times10^{-3}$ \\
$\langle {\cal O}^{\psi^\prime}[^3P_0^{(8)}]\rangle/$GeV$^5$  &
$(-2.08\pm2.28)\times10^{-3} $& 0 & 0

\\
\hline
$\chi^2/\mathrm{d.o.f}$ & --- & 0.033 & 0.051 \\
\hline $\langle {\cal O}^{\chi_{c0}}[^3P_0^{(1)}]\rangle/$GeV$^5$  &
$8.9\times10^{-2}$ & $8.9\times10^{-2}$ &
$8.9\times10^{-2}$ \\
$\langle {\cal O}^{\chi_{c0}}[^3S_1^{(8)}]\rangle/$GeV$^3$  & --- & $(2.14\pm 0.67)\times10^{-4}$ & $(1.69\pm0.9)\times10^{-4}$ \\
\hline
$\chi^2/\mathrm{d.o.f}$  & --- & 0.89 & 0.41 \\
\end{tabular}
\end{ruledtabular}
\end{center}
\end{table}

\newpage

\newpage

\begin{figure}[h]
\begin{center}
\includegraphics[width=0.95\textwidth]{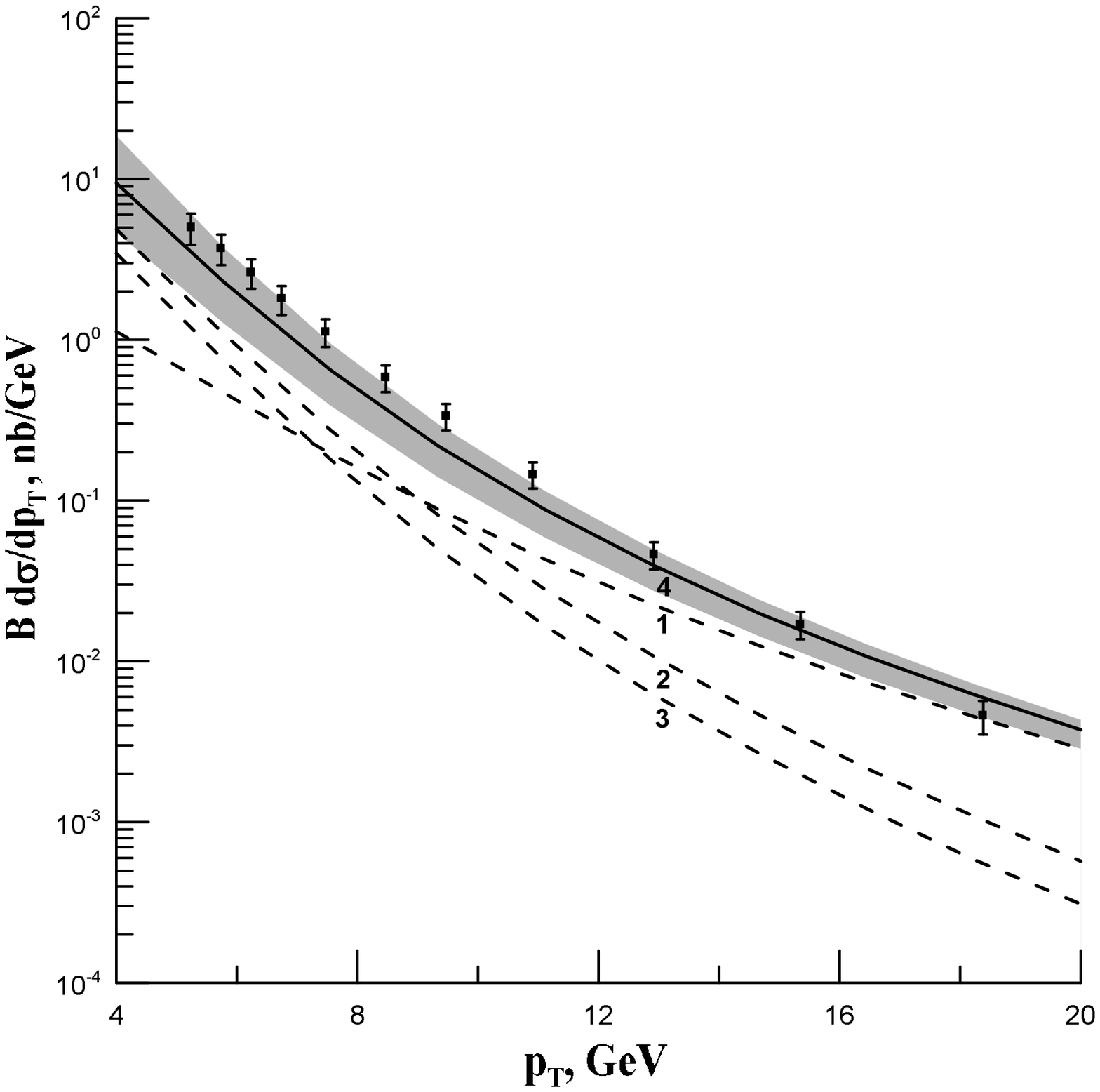}
\end{center}
\caption{\label{fig:1}Direct $J/\psi$ transverse momentum spectrum
from CDF Collaboration~\cite{CDFI}, $\sqrt{S}=1.8$ TeV,
$|\eta|<0.6$, (1) is the contribution of $[^3S_1^{(8)}]$ state, (2)
-- $[^1S_0^{(8)}]$, (3) -- $[^3S_1^{(1)}]$, (4) -- sum of their
all.}
\end{figure}

\newpage
\begin{figure}[h]
\begin{center}
\includegraphics[width=0.95\textwidth]{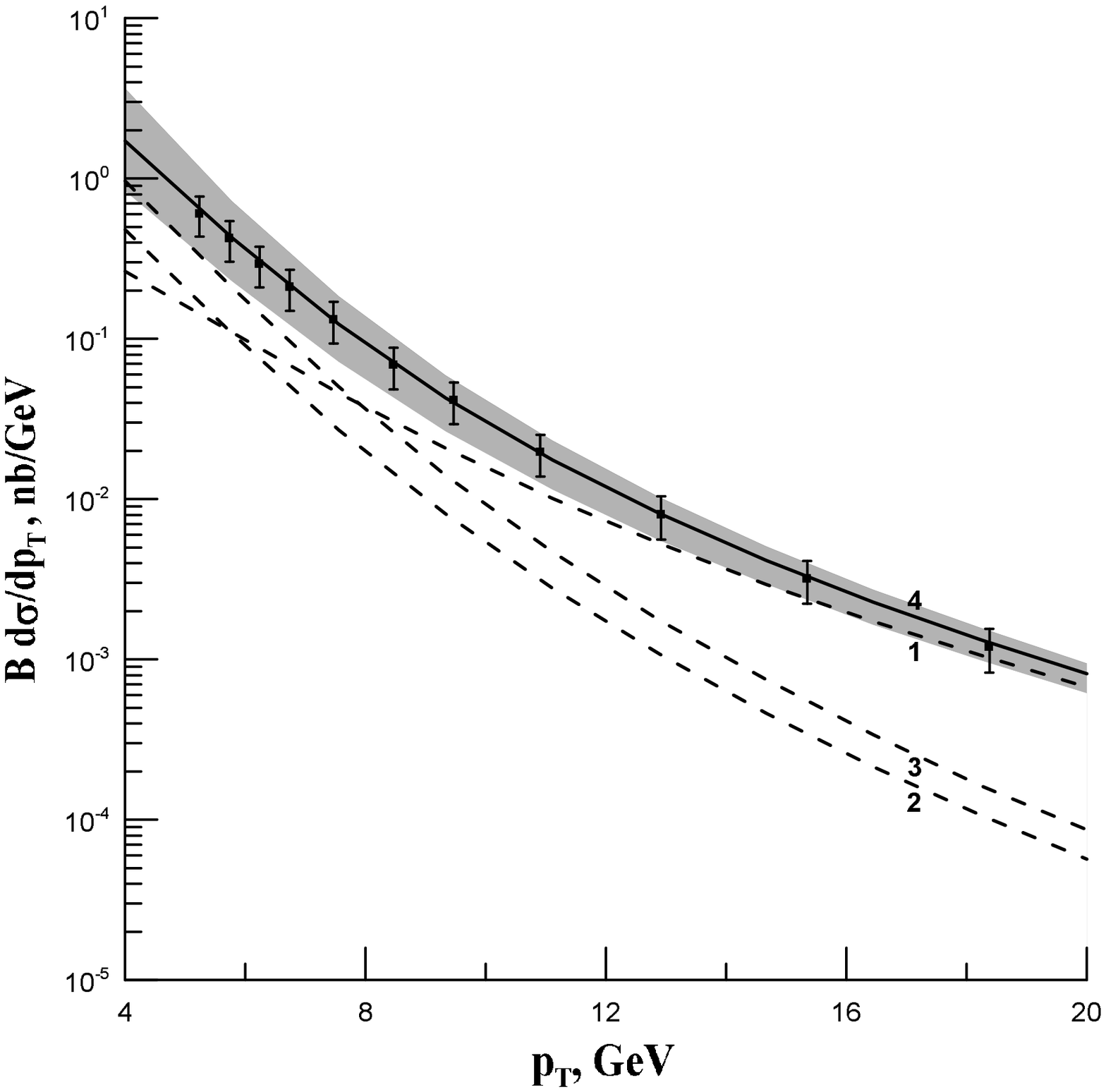}
\end{center}
\caption{\label{fig:1}$J/\psi$ transverse momentum spectrum from
$\psi'$ decays from CDF Collaboration~\cite{CDFI}, $\sqrt{S}=1.8$
TeV, $|\eta|<0.6$, (1) is the contribution of $[^3S_1^{(8)}]$ state,
(2) -- $[^3S_1^{(1)}]$, (3) -- $[^1S_0^{(8)}]$, (4) -- sum of their
all.}
\end{figure}
\newpage

\begin{figure}[h]
\begin{center}
\includegraphics[width=0.95\textwidth]{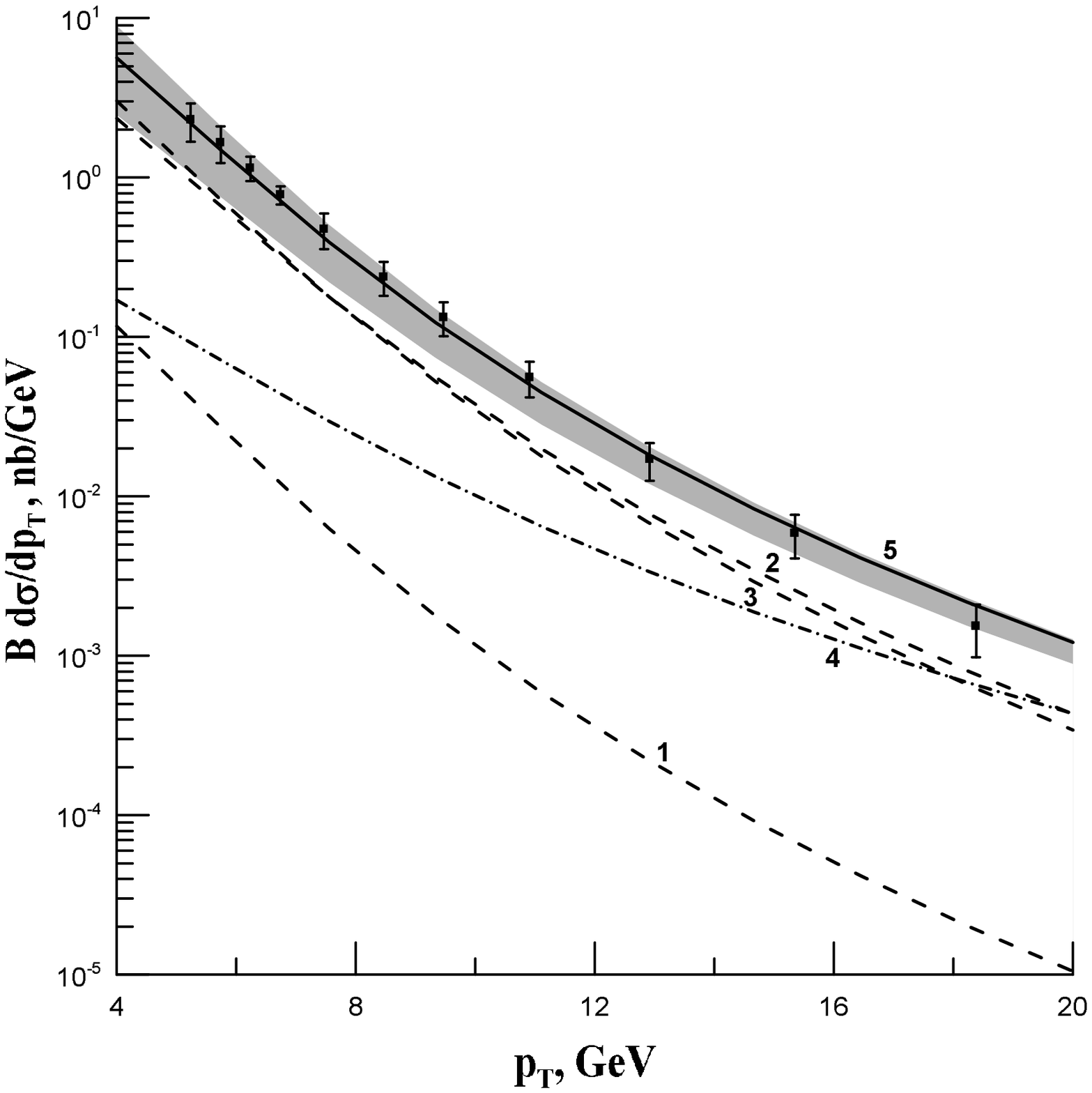}
\end{center}
\caption{\label{fig:3}$J/\psi$ transverse momentum spectrum from
$\chi_{cJ}$ decays from CDF Collaboration~\cite{CDFI},
$\sqrt{S}=1.8$ TeV, $|\eta|<0.6$, (1) is the contribution of
$[^3P_0^{(1)}]$ state, (2) -- $[^3P_1^{(1)}]$, (3) --
$[^3P_2^{(1)}]$, (4) -- $[^3S_1^{(8)}]$, (5) -- sum of their all.}
\end{figure}

\newpage
\begin{figure}[h]
\begin{center}
\includegraphics[width=0.95\textwidth]{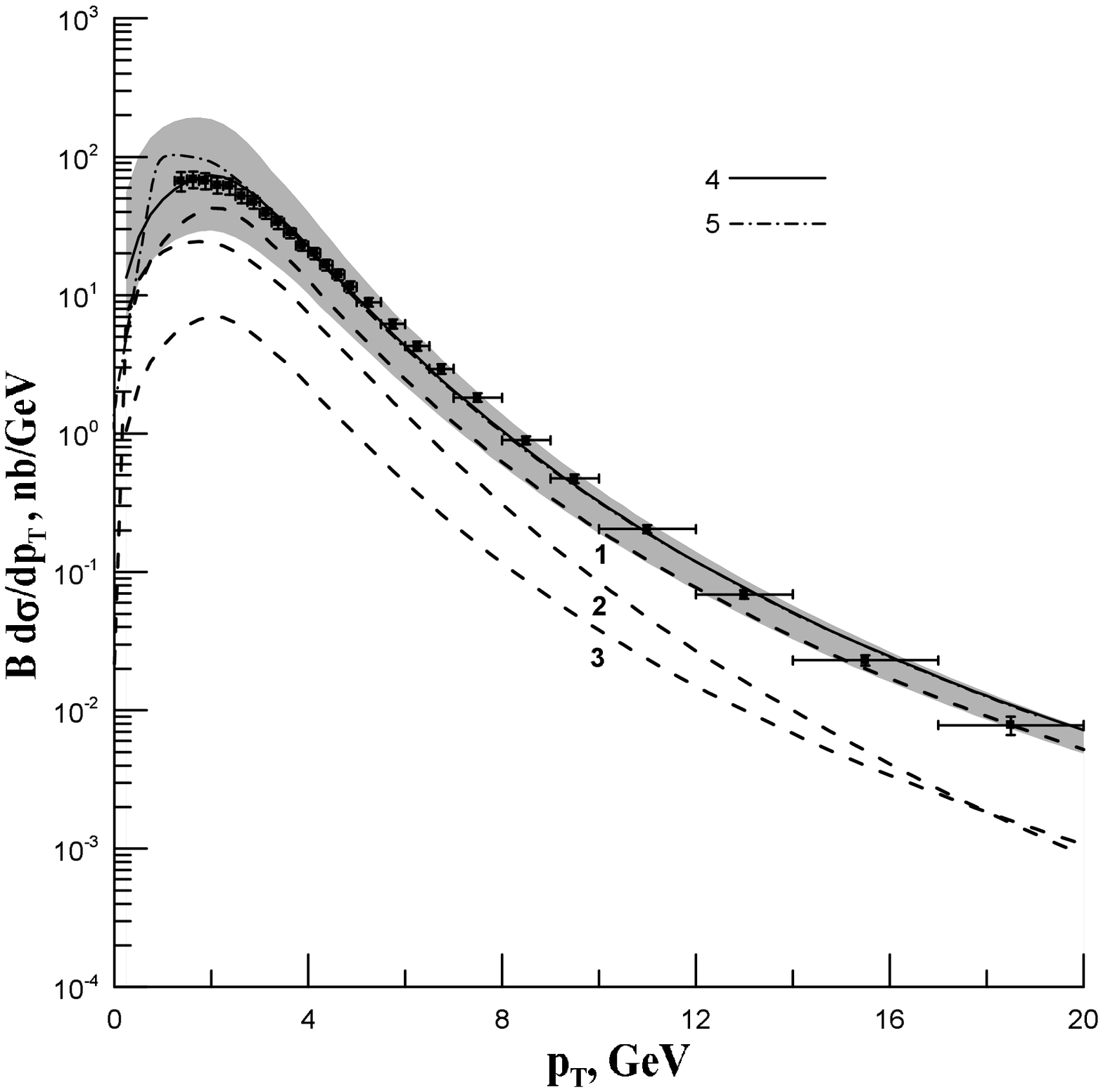}
\end{center}
\caption{\label{fig:4}Prompt $J/\psi$ transverse momentum spectrum
from CDF Collaboration~\cite{CDFII}, $\sqrt{S}=1.96$ TeV,
$|y|<0.6$, (1) is the direct production, (2) -- from $\chi_{cJ}$
decays, (3) -- from $\psi'$ decays, (4) -- sum of their all (KMR
unPDF), (5) -- sum of all contributions (Bl\"umlein unPDF).}
\end{figure}

\newpage
\begin{figure}[h]
\begin{center}
\includegraphics[width=0.8\textwidth]{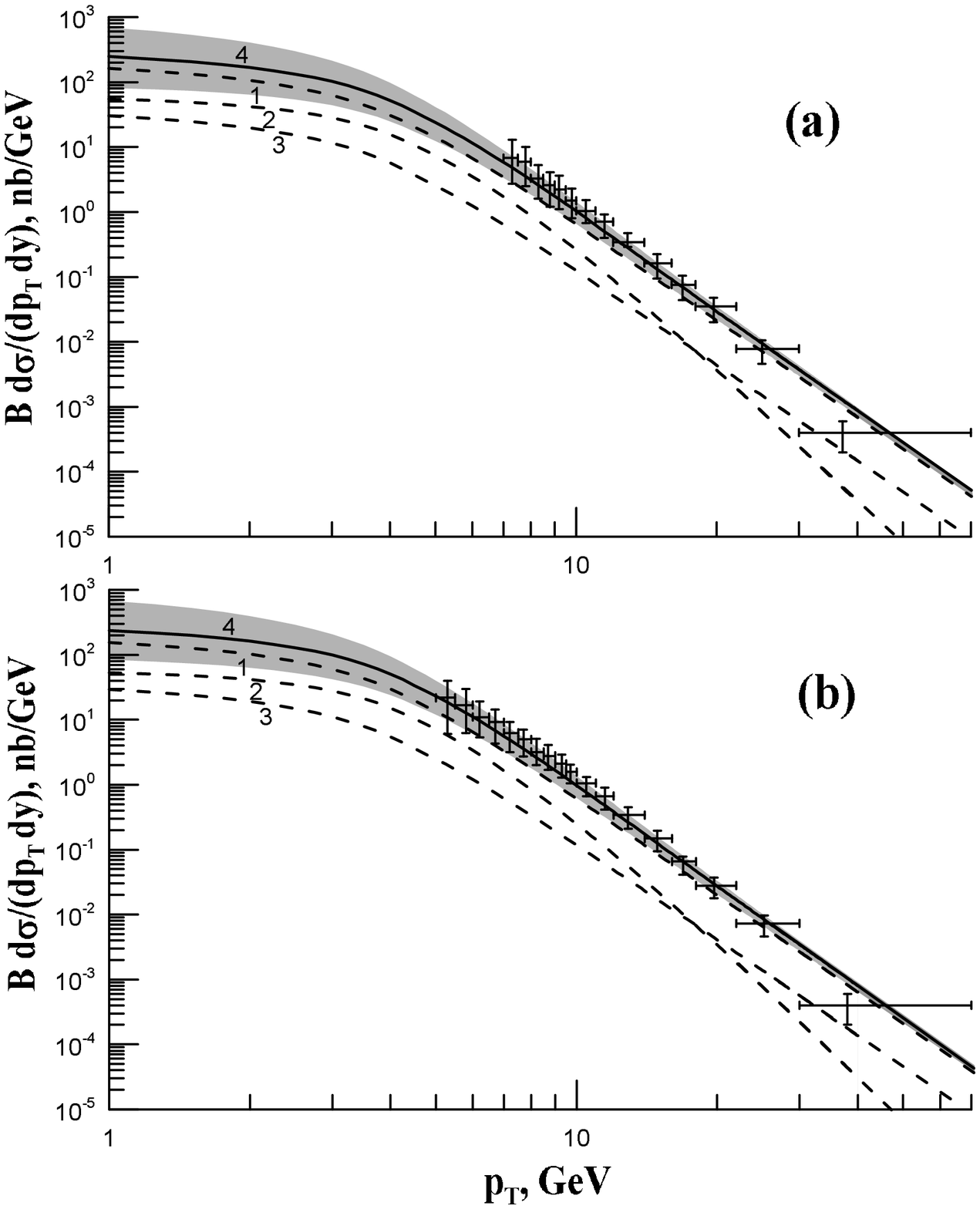}
\end{center}
\caption{\label{fig:5}Prompt $J/\psi$ transverse momentum spectrum
from ATLAS Collaboration~\cite{ATLASpsi}, $\sqrt{S}=7$ TeV, (1) is
the direct production, (2) -- from $\chi_{cJ}$ decays, (3) -- from
$\psi'$ decays, (4) -- sum of their all. For the different range in
the rapidity: (a)-- $|y|<0.75$, (b) -- $0.75<|y|<1.5$.}
\end{figure}
\newpage

\begin{figure}[h]
\begin{center}
\includegraphics[width=0.8\textwidth]{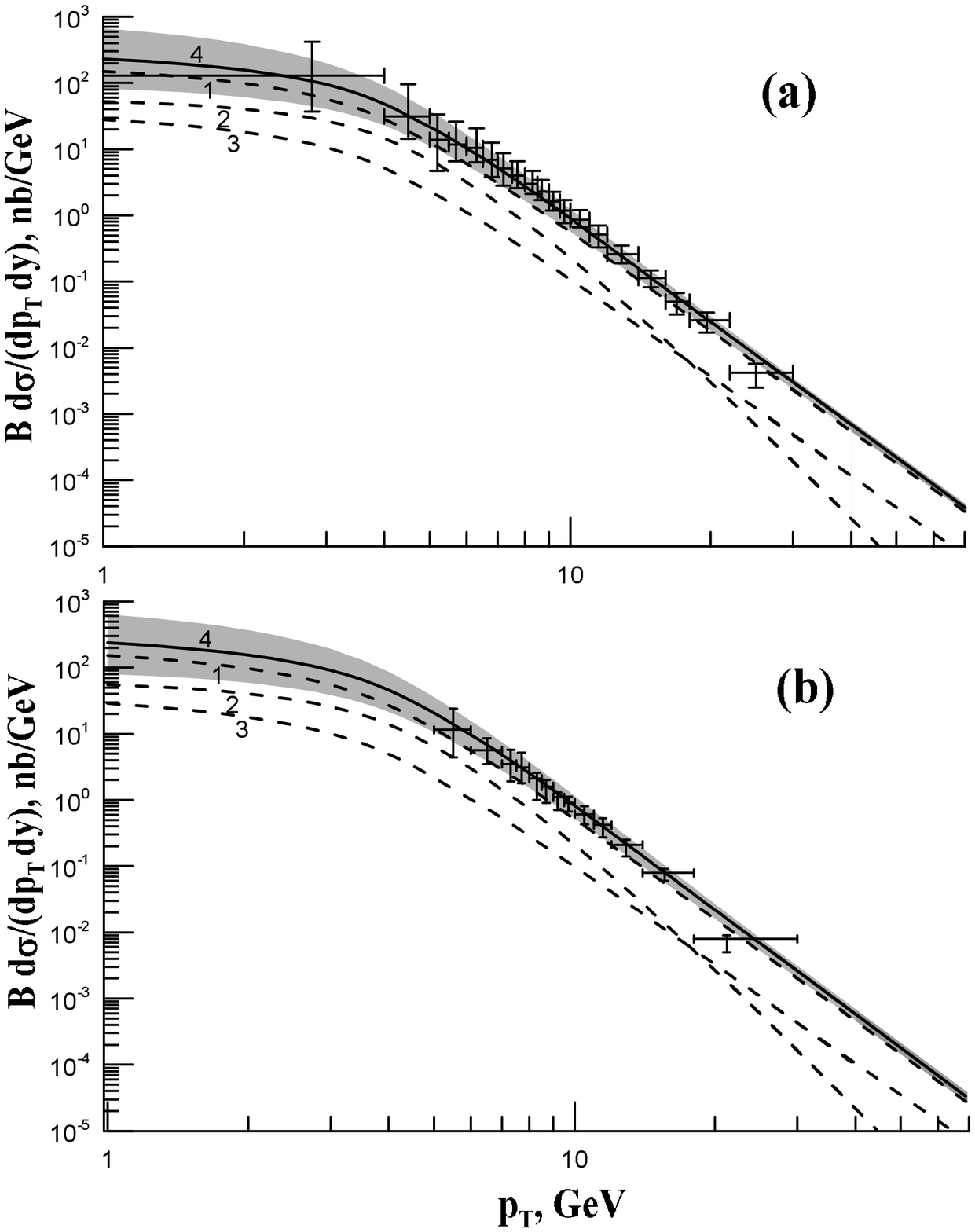}
\end{center}
\caption{\label{fig:6}Prompt $J/\psi$ transverse momentum spectrum
from ATLAS Collaboration~\cite{ATLASpsi}, $\sqrt{S}=7$ TeV, (1) is
the direct production, (2) -- from $\chi_{cJ}$ decays, (3) -- from
$\psi'$ decays, (4) -- sum of their all. For the different range in
the rapidity: a) -- $1.5<|y|<2.0$, (b) -- $2.0<|y|<2.4$  .}
\end{figure}

\newpage
\begin{figure}[h]
\begin{center}
\includegraphics[width=0.9\textwidth]{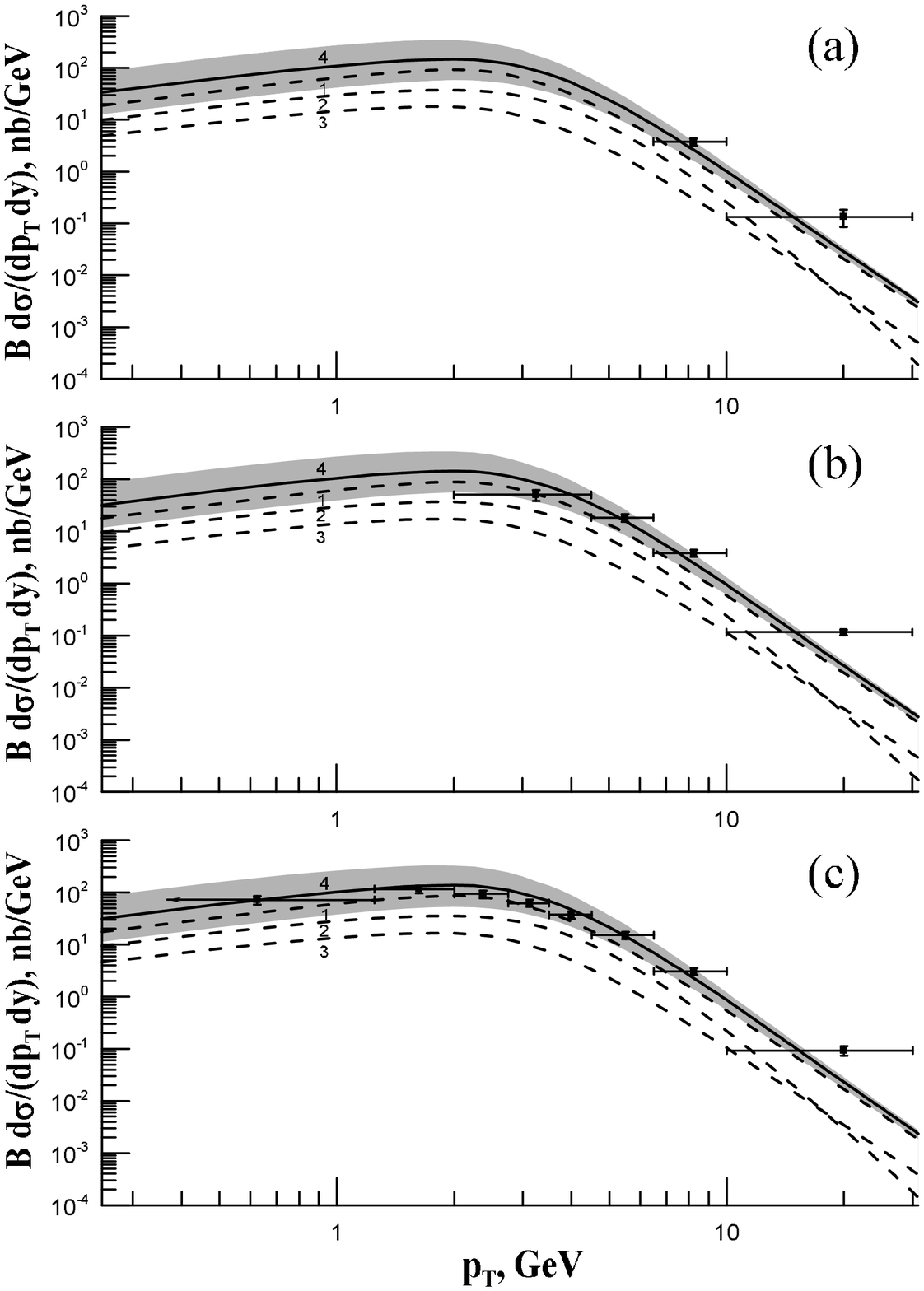}
\end{center}
\caption{\label{fig:7}Prompt $J/\psi$ transverse momentum spectrum
from CMS Collaboration~\cite{CMSpsi}, $\sqrt{S}=7$ TeV, (1) is the
direct production, (2) -- from $\chi_{cJ}$ decays, (3) -- from
$\psi'$ decays, (4) -- sum of their all. For the different range in
the rapidity: (a)-- $|y|<1.2$, (b) -- $1.2<|y|<1.6$, (c) --
$1.6<|y|<2.4$}
\end{figure}

\newpage
\begin{figure}[h]
\begin{center}
\includegraphics[width=0.9\textwidth]{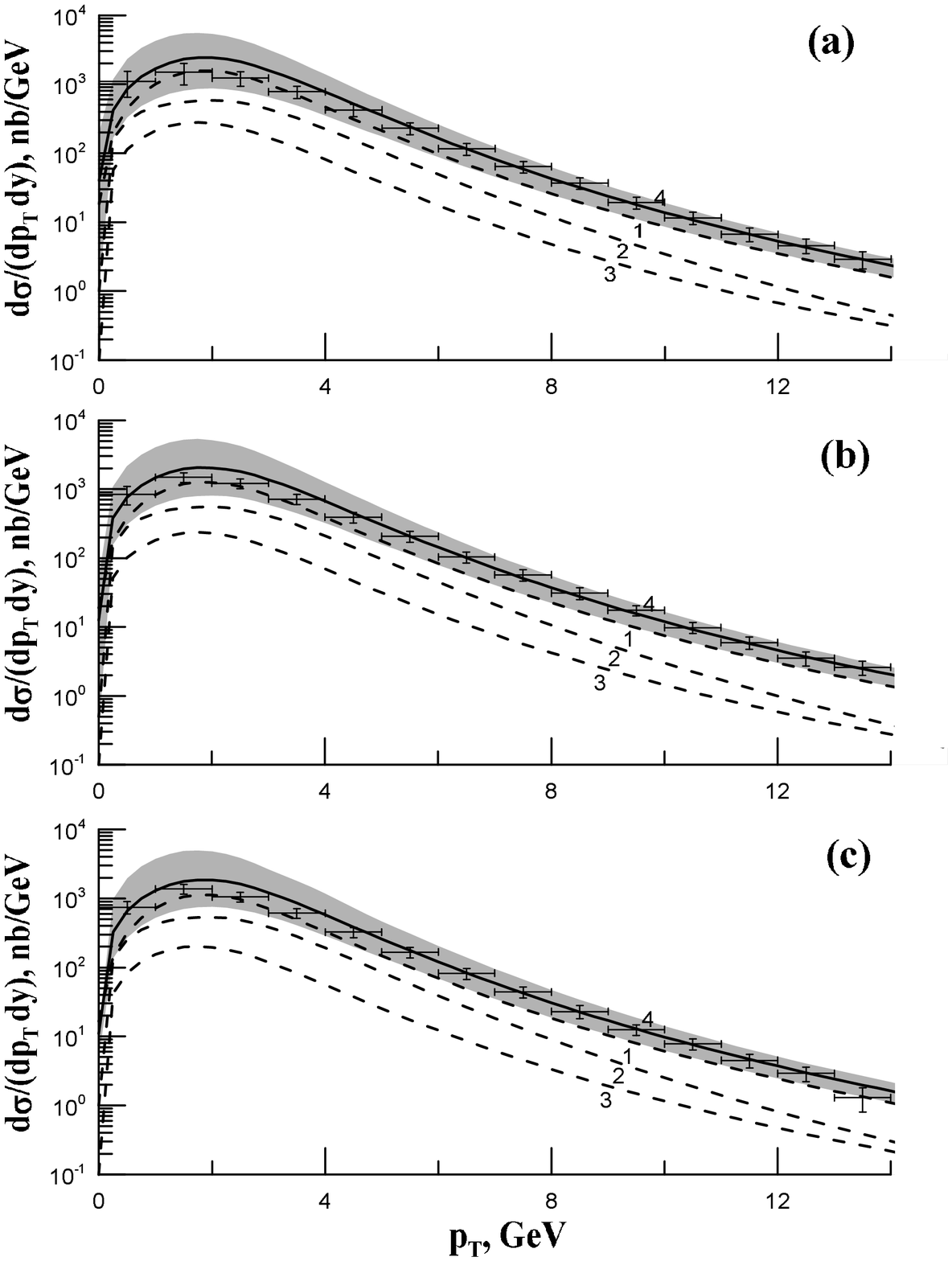}
\end{center}
\caption{\label{fig:8}Prompt $J/\psi$ transverse momentum spectrum
from LHCb Collaboration~\cite{LHCbpsi}, $\sqrt{S}=7$ TeV, (1) is the
direct production, (2) -- from $\chi_{cJ}$ decays, (3) -- from
$\psi'$ decays, (4) -- sum of their all. For the different range in
the rapidity: (a)- $2.0<|y|<2.5$, (b) - $2.5<|y|<3.0$, (c) -
$3.0<|y|<3.5$}
\end{figure}

\newpage
\begin{figure}[h]
\begin{center}
\includegraphics[width=0.9\textwidth]{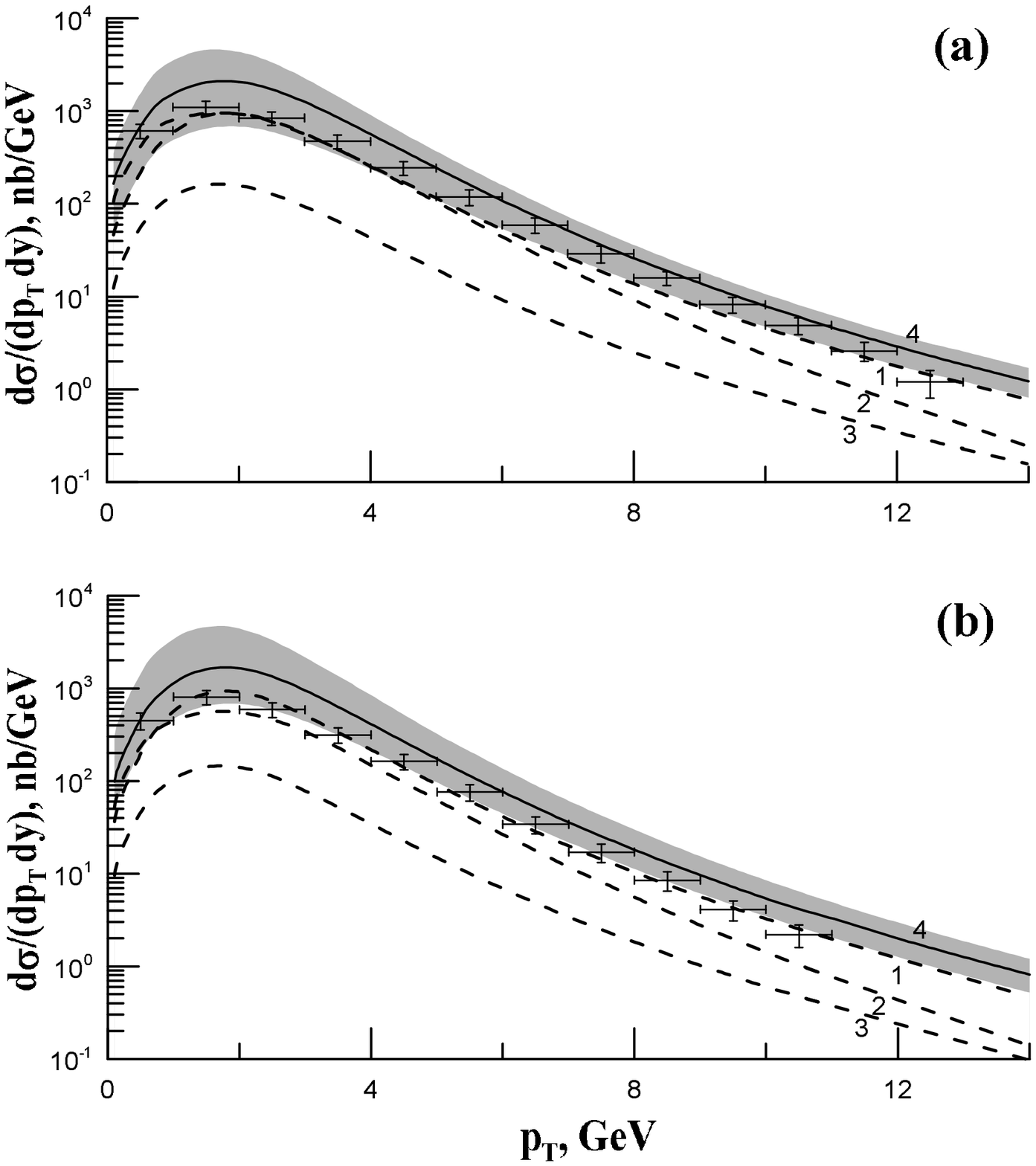}
\end{center}
\caption{\label{fig:9}Prompt $J/\psi$ transverse momentum spectrum
from LHCb Collaboration~\cite{LHCbpsi}, $\sqrt{S}=7$ TeV, (1) is the
direct production, (2) -- from $\chi_{cJ}$ decays, (3) -- from
$\psi'$ decays, (4) -- sum of their all. For the different range in
the rapidity: (a)-- $3.5<|y|<4.0$, (b) -- $4.0<|y|<4.5$}
\end{figure}

\newpage
\begin{figure}[h]
\begin{center}
\includegraphics[width=0.95\textwidth]{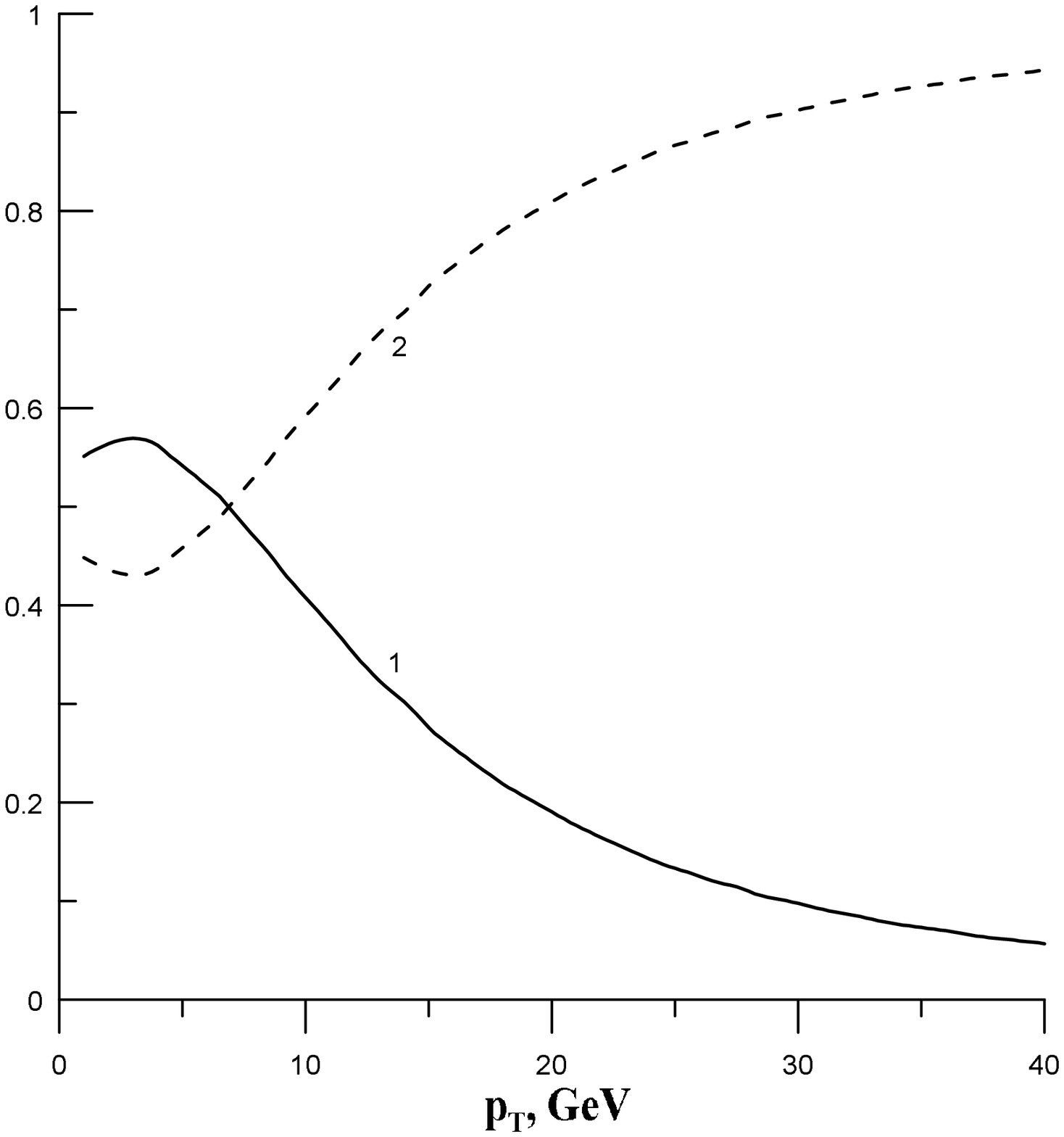}
\end{center}
\caption{\label{fig:10} The relative contributions of the
color-singlet (curve 1) and color-octet (curve 2) production
mechanisms to the prompt $J/\psi$ transverse momentum spectrum at
the $\sqrt{S}=7$ TeV, $1.5<|y|<2.0$.}
\end{figure}

\end{document}